\documentclass[aps,prd,twocolumn,groupedaddress,nofootinbib,floatfix,superscriptaddress]{revtex4-1}

\usepackage{amsmath,amssymb,amsfonts,graphicx}
\usepackage{aas_macro}

\usepackage[dvipsnames]{xcolor}

\usepackage[colorlinks]{hyperref}
\hypersetup{
    colorlinks=true,
    linkcolor=NavyBlue,
    filecolor=NavyBlue,      
    urlcolor=NavyBlue,
    citecolor=NavyBlue,
    pdfpagemode=FullScreen
}

\begin{document}

\title[AMS-02 Be data and its implication for CR transport]{
AMS-02 beryllium data and its implication for cosmic ray transport}

\author{\href{https://orcid.org/0000-0002-6023-5253}{Carmelo Evoli}}
\email[]{carmelo.evoli@gssi.it}
\affiliation{Gran Sasso Science Institute (GSSI), Via M. Iacobucci 2, 67100 L'Aquila, Italy}
\affiliation{INFN-Laboratori Nazionali del Gran Sasso, Via G. Acitelli 22, Assergi (AQ), Italy}

\author{\href{https://orcid.org/0000-0002-5014-4817}{Giovanni Morlino}}
\email[]{giovanni.morlino@inaf.it}
\affiliation{INAF-Osservatorio Astrofisico di Arcetri, L.go E. Fermi 5, 50125 Firenze, Italy}

\author{\href{https://orcid.org/0000-0003-2480-599X}{Pasquale Blasi}}
\email[]{pasquale.blasi@gssi.it}
\affiliation{Gran Sasso Science Institute (GSSI), Via M. Iacobucci 2, 67100 L'Aquila, Italy}
\affiliation{INFN-Laboratori Nazionali del Gran Sasso, Via G. Acitelli 22, Assergi (AQ), Italy}

\author{\href{https://orcid.org/0000-0003-0161-5923}{Roberto Aloisio}}
\email[]{roberto.aloisio@gssi.it}
\affiliation{Gran Sasso Science Institute (GSSI), Via M. Iacobucci 2, 67100 L'Aquila, Italy}
\affiliation{INFN-Laboratori Nazionali del Gran Sasso, Via G. Acitelli 22, Assergi (AQ), Italy}

\date{\today}

\begin{abstract}
The flux of unstable secondary cosmic ray nuclei, produced by spallation processes in the interstellar medium, can be used to constrain the residence time of cosmic rays inside the Galaxy. Among them, $^{10}$Be is especially useful because of its relatively long half-life of 1.39 Myr. In the framework of the diffusive halo model we describe cosmic ray transport taking into account all relevant interaction channels and accounting for the decay of unstable secondary nuclei. We then compare our results with the data collected by the Alpha Magnetic Spectrometer (AMS-02) on board the International Space Station for the flux ratios Be/C, B/C, Be/O, B/O, C/O and Be/B as well as C, N and O absolute fluxes. These measurements, and especially the Be/B ratio, allow us to single out the flux of $^{10}$Be and infer a best fit propagation time of CRs in the Galaxy. Our results show that, if the cross sections for the production of secondary elements through spallation are taken at face value, AMS-02 measurements are compatible with the standard picture based on CR diffusion in a halo of size $H \gtrsim 5$ kpc. Taking into account the uncertainties in the cross sections, this conclusion becomes less reliable, although still compatible with the standard picture. Implications of our findings for alternative models of CR transport are discussed. 
\end{abstract}

\maketitle

%%%%%%%%%%%%%%%%%% BODY OF PAPER %%%%%%%%%%%%%%%%%%

\section{Introduction}

The transport of cosmic rays (CRs) in the Galaxy is a complex phenomenon ruled by microphysical processes affecting the large scale behaviour of charged particles in the pervading magnetic fields of the interstellar medium (ISM). Such small scale complexity is typically averaged out in such a way that simplified equations are found that describe the spatial dependence of the CR spectrum throughout the Galaxy. The most common approach to CR transport is based on the so-called diffusion-convection equation, sometimes with the additional assumption that the region where sources are located and where CR interactions occur has a scale height that is much smaller than the size of the magnetized underdense Galactic halo. The spectrum of CRs observed at the Earth is then a convolution of the source spectrum and the confinement time of CRs in the Galaxy. The interactions suffered by CRs in their journey through the Galaxy is described in terms of the total mass per unit surface, the so-called grammage. This quantity can be measured by using the ratio of fluxes of secondary-to-primary nuclei, such as the boron (B) to carbon (C) ratio. In the context of the standard diffusion-convection model, these ratios are typically proportional to $H/D(E)$ at high energies, so that precious information about CR transport can be gathered through their measurement. However, the combination of the halo size $H$ and the diffusion coefficient $D(E)$ that determines the secondary-to-primary ratios leaves the confinement time $H^{2}/D(E)$ weakly constrained: the same ratios could be obtained by assuming small halo size and correspondingly small diffusion coefficient or assuming that both quantities are larger by the same amount. On the other hand, the propagation of primary leptons from the sources to the Earth is regulated by the balance between confinement time and radiative losses, hence lepton transport requires the knowledge of the confinement time $\sim H^{2}/D(E)$. 

The measured secondary-to-primary ratios of nuclei are all decreasing functions of energy at $E\gtrsim 10$ GeV/n, thereby confirming the theoretical expectation that CR transport is mainly diffusive and that the grammage associated to diffusive propagation is also a decreasing function of energy in the same energy range. Some features in the spectra of primary nuclei~\cite{PAMELA-phe,AMS02-H,AMS02-he,AMS02-heco} and a peculiar trend in the secondary-to-primary ratios recently measured by the AMS-02 collaboration~\cite{AMS02-bc,AMS02-libeb} stimulated some discussion about the possibility that these features may reflect different regimes in the diffusive transport of CRs in the Galaxy~\cite{Tomassetti2012,Blasi2012,Aloisio2013,Aloisio2015,Evoli2018b}.

On the other hand, this picture of CR transport appears to be challenged by observations of the fluxes of positrons and antiprotons, also produced as secondary products of CR inelastic interactions~\cite{Evoli2015}. In fact the flux ratios $e^{+}/(e^{+} + e^{-})$ and $\bar{p}/p$, that are both expected to decrease with energy, are observed to increase and be roughly constant with energy respectively~\cite{PAMELA-posfraction,PAMELA-antip-p,AMS02-posfraction,AMS02-antip}. It should be emphasized that a contribution to the positron flux at the Earth is actually expected based on models of particle escape from pulsars (see \cite{Amato2018} for a recent review). This contribution has been shown to lead to an increasing positron ratio.
Moreover, an improved knowledge of the cross section for production of antiprotons, a somewhat updated choice of propagation parameters~\cite{Reinert2018,Boudaud2019} and the introduction of physical processes such as first order reacceleration and source grammage~\cite{Bresci2019}, make the predicted shape of the $\bar{p}/p$ more similar to the observed one.

Nevertheless, the occurrence that the shape of the spectra of positrons, antiprotons and protons is very similar justifies some doubts concerning the standard picture of CR transport illustrated above. In fact several authors have advocated the idea that positrons and antiprotons are purely secondary products of CR interactions \cite{Blum2013,Katz2010,Lipari2017}, with no need for alternative sources or exotic models, provided the standard interpretation of the grammage is deeply changed. In order to accommodate the energy decrease of the grammage as inferred from the B/C ratio (and other secondary-to-primary ratios) it was suggested, following the idea of nested leaky-box put forward in Ref.~\cite{Cowsik2016}, that sources may be surrounded by regions of enhanced grammage, the so-called cocoons~\cite{Lipari2017}. These regions would dominate CR grammage up to $\sim$TeV/n, while at higher energies the grammage would be accumulated mainly throughout the ISM, and in an energy independent manner. None of these assumptions, at this point, is much more than just a working hypothesis. 

However some recent works have showed how the streaming of CRs away from their sources and the large CR densities in such regions may induce instabilities that self-confine CRs for times that largely exceed the ones that may be naively expected \cite{Ptuskin2008,Malkov2013,Dangelo2016}. 
{Depending on the ISM conditions around the sources (especially the neutral fraction~\cite{Nava2016,Nava2019}) this may result in enhanced near source grammage, conceptually similar to the cocoons mentioned above.}
Although a clear picture of all relevant elements of this problem is not yet available, it is interesting that in a rather independent way, in the last few years several pieces of observations, typically in the form of extended gamma ray halos around pulsar wind nebulae, supernova remnants and star clusters, led to estimates of the diffusion coefficient in these near source regions up to $\sim$100 times smaller than typically inferred for the ISM~\cite{Hanabata2014,Abeysekara2017,Aharonian2019}.

In the alternative approaches discussed above~\cite{Cowsik2016,Lipari2017}, the spectral similarity between positrons and both protons and antiprotons forces one to require that the observed positron spectrum is not appreciably affected by radiative losses, again in striking contradiction with the standard scenario, in which energy losses dominate leptons' transport for energies above few GeV. As pointed out in Ref.~\cite{Lipari2017}, this requirement implies that the escape time of CRs from the Galaxy in the $\sim$10~GeV energy range be of order a few million years rather than the typical value of $\sim$100 Myr, as inferred in the standard picture. In order to support this finding, it is typically argued that the few existing measurements of the $^{10}$Be/$^{9}$Be at low energies suggest a much shorter confinement time than in the standard model~\cite{Garcia-Munoz1977,Ahlen2000,Yanasak2001}, in which case it may in fact be reasonable that positrons with energy $\lesssim 1$ TeV may be little affected by radiative losses. This conclusion is usually based upon the adoption of some variation of the so-called leaky box model, that is known to be unfit to the description of unstable isotopes, such as $^{10}$Be~\cite{Ptuskin1998}. The main reason for such limitation is that at low energies the decay of $^{10}$Be takes place inside the disc of the Galaxy, in striking contradiction with the basic assumption of the leaky box model. 

Even in the case of diffusion-advection approaches with an infinitely thin disc this situation would be ill described. The latter would be an appropriate approach at higher energies, where the Lorentz boosted decay time may exceed the time for escaping the Galactic disc, but until recently no measurement existed of the decaying isotopes at $E\gtrsim 10$ GeV/n, and the existing measurements at lower energies are affected by substantial systematic uncertainties~\cite{Webber1979,Garcia-Munoz1981,Connell1998,Hams2004,Lukasiak1999}.

Nowadays, the unprecedented quality of the data collected by the AMS-02 mission, onboard the International Space Station, is providing extremely detailed information on the fluxes of CRs, both of primary and secondary nature, up to energies of order $\sim$TeV. In particular AMS-02 recently published the observed fluxes of secondary CRs such as Lithium, Beryllium and Boron~\cite{AMS02-libeb}. Although not designed to carry out an isotopic analysis of unstable elements, AMS-02 measured the total spectrum of Beryllium and the energy dependence of the Be/B ratio, which contains precious information about the confinement time, as we discuss below and as first proposed in \cite{Daniel1966}. In the absence of decays of $^{10}$Be, this ratio above $\sim 10$ GeV should be a slightly decreasing function of energy, as a result of the mildly larger cross section of Boron spallation which reduces the denominator of the ratio.  On the other hand, if at a given energy an appreciable fraction of $^{10}$Be may decay, the total flux of Beryllium decreases. Moreover, the decays of $^{10}$Be mainly result in the production of Boron nuclei. Both these effects invert the expected trend, so that the Be/B ratio can now be expected to be an increasing function of energy, to an extent which depends on the fraction of $^{10}$Be nuclei that decay, which in turn carries information about the confinement time in the Galaxy. In this article we discuss in detail the results of our investigation of this effect. 

The article is organised as follows, in Sec.~\ref{sec:Model} we introduce the formalism used to describe CR transport in the Galaxy, both for primary nuclei and for secondary stable and unstable nuclei. The results of our calculations are discussed in Sec.~\ref{sec:Results} in connection with the AMS-02 data. The conclusions and an outlook for future measurements are then presented in Sec.~\ref{sec:Conclusions}.

\section{Model}
\label{sec:Model}

The theoretical approach that we adopted to describe CR propagation is based on the diffusive halo model and is a modified version of the weighted slab technique already introduced in \cite{Jones2001,Aloisio2013,Aloisio2015,Evoli2019}. Within this approach, the CR sources are assumed to be located in a thin disc with half-width $h\ll H$, where $H$ is the half thickness of the Galactic halo. The ISM gas that acts as target for CR interactions is also assumed to be confined inside the thin disc, with a surface density is $\mu=2.3$ mg/cm$^2$ \cite{Ferriere2001}. The weighted slab technique has been generalized here to include two important effects: 1) the decay of unstable nuclei; 2) the contribution to stable nuclei (such as $^{10}$B) from the decay of unstable isotopes (such as $^{10}$Be). 

The adoption of the weighted slab model is justified for the description of the decay of $^{10}$Be if such decays take place outside the thickness $h$ of the disc. It is easy to check what are the constraints that this condition imposes on the energy per nucleon of the decaying nucleus. The relevant timescales for CR transport in the disc are the diffusion timescale $h^{2}/D(R)$ and the advection timescale $h/v_A$, where $v_{A}$ is the Alfv\'en speed. In order for the model to be applicable we require that the decay of $^{10}$Be takes place in the halo
$$\gamma \tau_d \gg \rm Min\left[ \frac{h^2}{2 D}, \frac{h}{v_{A}} \right],$$ 
where $\tau_d = t_{1/2} / \ln{2} \sim 2$~Myr is the timescale for the radioactive decay of $^{10}$Be, $\gamma$ is the Lorentz factor and $D(R)$ is the rigidity-dependent diffusion coefficient. As in~\cite{Evoli2019}, we assume a diffusion coefficient that is spatially constant and only dependent upon particles rigidity $R$:
\begin{equation}
D(R) = 2 v_A H + \beta D_0 \frac{(R/{\rm GV})^\delta}{[1+(R/R_b)^{\Delta \delta /s}]^s},
\label{diffcoe} 
\end{equation}
where $D_0$ and $\delta$ are parameters that are fitted to the data, mainly the B/C and B/O ratios as functions of energy. The other parameters $s$, $\Delta \delta$ and $R_{b}$ are fixed from observations of primary nuclei~\cite{Evoli2019}:  $s = 0.1$, $\Delta \delta = 0.2$, $R_b = 312$~GV. The functional form in Eq.~\eqref{diffcoe}, also used in Ref.~\cite{Evoli2019}, is inspired to (but not limited to) the models in which the diffusion coefficient is self-generated by propagating CRs~\cite{Blasi2012,Aloisio2013,Aloisio2015}. 
The plateau at low energies, where advection dominates transport, was found in self-generated models in Ref.~\cite{Recchia2016b}. 

Rather than determining $v_{A}$ from physical quantities, some of which are very poorly known in the halo, we fit the value of $v_{A}$ to the existing data on the fluxes of both primary and secondary nuclei. 
\begin{figure}[t]
\begin{center}
\includegraphics[width=\columnwidth]{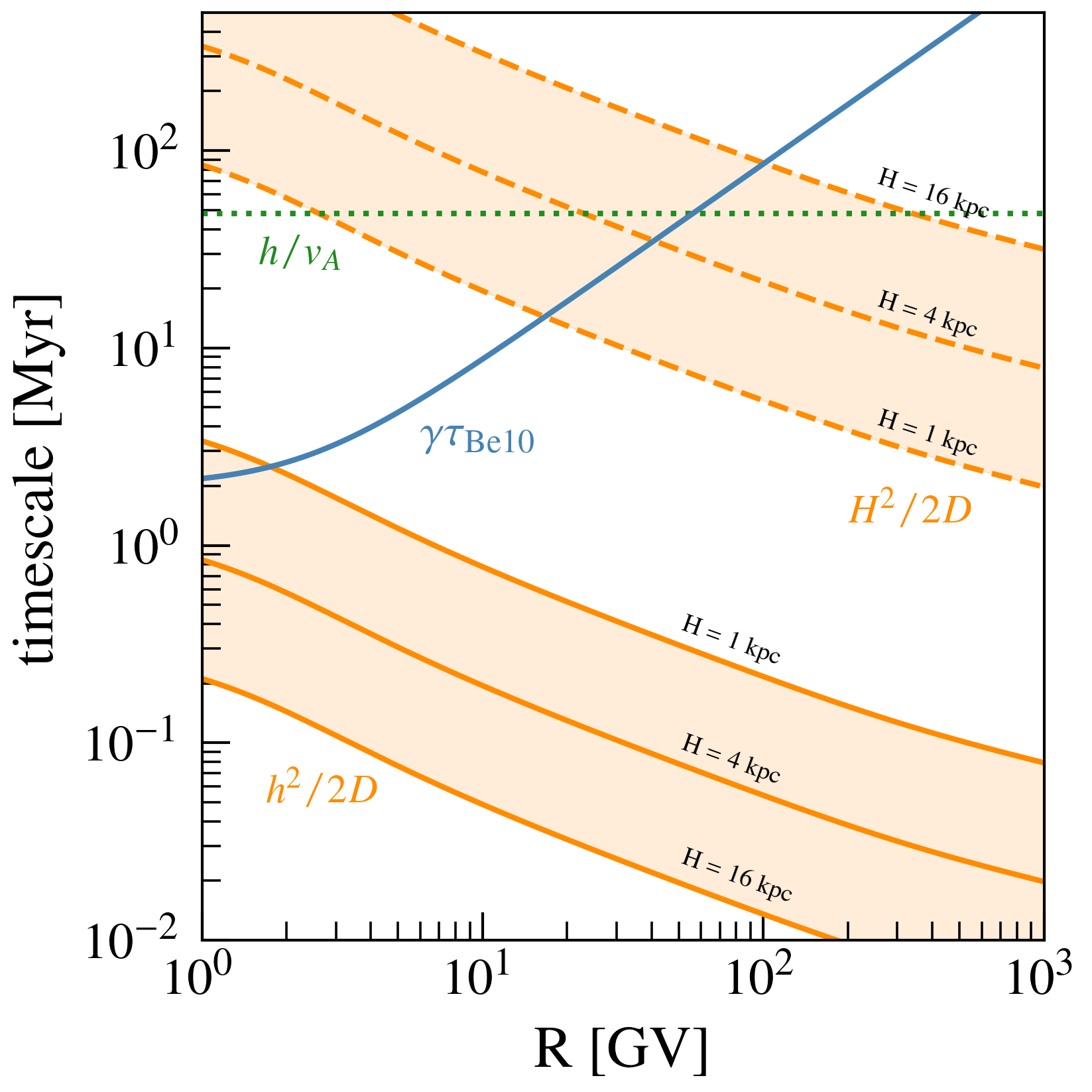}
\end{center}
\caption{Diffusion time scale in the disk $h$ (solid orange lines) and in the halo $H$ (dashed orange lines) for three different values of the halo size. We also show the Lorentz boosted decay time of $^{10}$Be (blue solid line) and the advection timescale to exit the disc (green dotted line).}
\label{fig:be10timescales}
\end{figure}

In Fig.~\ref{fig:be10timescales} we illustrate the limits of validity of the assumption of $^{10}$Be decay outside the thin disc. The dependence of the results on the size $H$ of the halo is due to the fact that the secondary-to-primary ratios approximately fix the ratio of the normalization of the diffusion coefficient and the halo size $H$. This implies that larger halos require correspondingly larger diffusion coefficients. From Fig.~\ref{fig:be10timescales} it is clear that for $H\gtrsim 2$ kpc the Lorentz boosted decay time is appreciably longer than the diffusion time of the same nuclei in the Galactic disc. Even for $H\sim 1$ kpc, this condition is well satisfied for rigidity $\gtrsim$ few GV. The advection time is irrelevant for transport on spatial scales $h \sim 150$~pc, being always much longer than the diffusion timescale for values of $v_A \sim 10$~km/s. 

It might be argued that the validity of the assumption of $^{10}$Be decay in the halo also depends upon the ansatz that the diffusion coefficient in the disc is the same as in the halo. This is partially true. On the other hand, if to consider the microphysics of particle transport, the Galactic disc is a rather hostile environment for CR scattering, because of severe ion-neutral damping of Alfv\'en waves for CR energies below $\sim 100$ GeV \cite[see][and references therein for a recent review]{Amato2018}. This would imply an even larger diffusion coefficient in the disc, thereby making the condition of $^{10}$Be decay in the halo easier to fulfil.  

The decay time of $^{10}$Be becomes longer than the escape time from the Galactic halo for rigidity above 10-100 GV, depending on the size $H$ of the halo, which is exactly the reason why the measurement of the flux of this isotope is sensitive to the parameter $H$. 

The transport equation describing the propagation of both stable and unstable nuclei in the context of the modified weighted slab approach reads:
\begin{multline}\label{eq:slab}
-\frac{\partial}{\partial z} \left[ D_{a} \frac{\partial f_a }{\partial z} \right] 
+ v_A \frac{\partial f_a}{\partial z}
- \frac{dv_A}{dz} \frac{p}{3} \frac{\partial f_a}{\partial p} 
\\
+ \frac{1}{p^{2}} \frac{\partial}{\partial p} \left[ p^{2} \left(\frac{dp}{dt}\right)_{a,\rm ion} f_a \right] 
+ \frac{\mu v(p) \sigma_a}{m} \delta(z) f_a 
+ \frac{f_a}{\hat\tau_{d,a}}
\\ 
= 2 h_d q_{0,a}(p) \delta(z) 
+ \sum_{a' > a} \frac{\mu\, v(p) \sigma_{a' \to a}}{m}\delta(z) f_{a'}
+ \sum_{a' > a} \frac{f_{a'}}{\hat\tau_{d,a'}},
\end{multline}
where $f_a (p,z)$ is the distribution function of specie $a$ in phase space, $v(p)=\beta(p) c$ is the particles' velocity, and $\mu$ is the surface density of the disk. The quantities $\hat \tau_{d, a}=\gamma \tau_{d,a}$ define the Lorentz boosted decay times of unstable elements. 

The second term on the LHS of Eq.~\ref{eq:slab} accounts for particle advection with velocity $v_A$. In the simple scenario adopted here, where the advection speed is constant in $z$, one has $dv_A/dz = 2v_{A} \delta(z)$ \cite{Evoli2019}.
The injection of primary CR nuclei of type $a$ occurs in the infinitely thin disc and is described through the function $q_{0,a}(p)$, assumed to be a power law in momentum with a slope $\gamma_{\rm inj}$ that depends slightly on the type of primary nucleus, as discussed in Ref.~\cite{Evoli2019}.
The second term in the RHS of Eq.~\ref{eq:slab} takes into account the production of secondary CRs through spallation processes while the third term accounts for the production of secondary CRs through radioactive decays of secondaries, such as $^{10}$B produced by the decay of $^{10}$Be. Notice that this latter term behaves as an injection term that however is not spatially relegated in the thin disc, which implies some technical difficulties, described below. 
Spallation cross sections are computed using the parametric fits provided by~\cite{Evoli2019} for LiBeB production from major primary channels and we adopted the approach described in~\cite{Evoli2018} to compute all the other secondary production cross sections.
While in the limit $\tau_{d,a} \to \infty$ Eq.~\ref{eq:slab} reduces to the standard transport equation for stable nuclei, the case including the decay of unstable isotopes requires some care, both because of the decay itself and because of the fact that some radioactive decays (such as $^{10}$Be) result in the spatially distributed injection of stable nuclei (for instance $^{10}$B \cite{Morlino2019}).
Finally we account for the effect of solar modulation by using the force field approximation~\cite{Gleeson1968} with a Fisk potential $\phi$ treated as one of the fitting parameters. 

\subsection{Unstable nuclei}

The solution of the transport equation for unstable elements can be found by using a procedure that is very similar to the one previously illustrated for stable nuclei~\cite{Aloisio2013,Aloisio2015,Evoli2019}. Let us first consider Eq.~\ref{eq:slab} for $z\neq 0$, where all terms proportional to $\delta(z)$ disappear and the equation reduces to:
\begin{equation}\label{eq:halo}
-\frac{\partial}{\partial z} \left[D_{a}(p) \frac{\partial f_{a}}{\partial z}\right] + v_A \frac{\partial f_{a}}{\partial z}
+ \frac{f_a}{\tau_{d,a}} = 0.
\end{equation}

The solution of this equation is readily found to be in the form:
\begin{equation}\label{eq:stable_halo}
f_a = A e^{\alpha_+ z} + B e^{\alpha_- z} 
\end{equation}
where $\alpha_{\pm}$ are the solutions of the second order algebraic equation $D_a \alpha^2 - v_A \alpha -1/\tau_{d,a}=0$: 
\begin{equation}
\alpha_{\pm} = \frac{v_A}{2 D_a} \left[ 1 \pm \sqrt{1+ \frac{4 D_a}{v_A^2 \tau_{d,a}} }\right] 
\equiv \frac{v_A}{2 D_a} \left[ 1 \pm \Delta_a \right]\,.
\end{equation}
Here we have introduced the dimensionless quantity $\Delta_a$ that can be written more conveniently as a function of the timescales involved in the propagation process, namely
\begin{equation}
\Delta_a = \sqrt{1+ 2 \tau_{\rm adv}^2/ \left(\tau_{{\rm diff},a} \, \tau_{d,a} \right) } \,
\end{equation}
where $\tau_{{\rm diff},a} = H^2/(2D_a)$ and $\tau_{\rm adv}= H/v_A$. 

In the limit of stable nuclei, $\tau_d \rightarrow \infty$, $\Delta \rightarrow 1$ and the solution in Eq.~\eqref{eq:stable_halo} reduces to the one found in Ref.~\cite{Evoli2019}.

\begin{figure}[t]
\begin{center}
\includegraphics[width=\columnwidth]{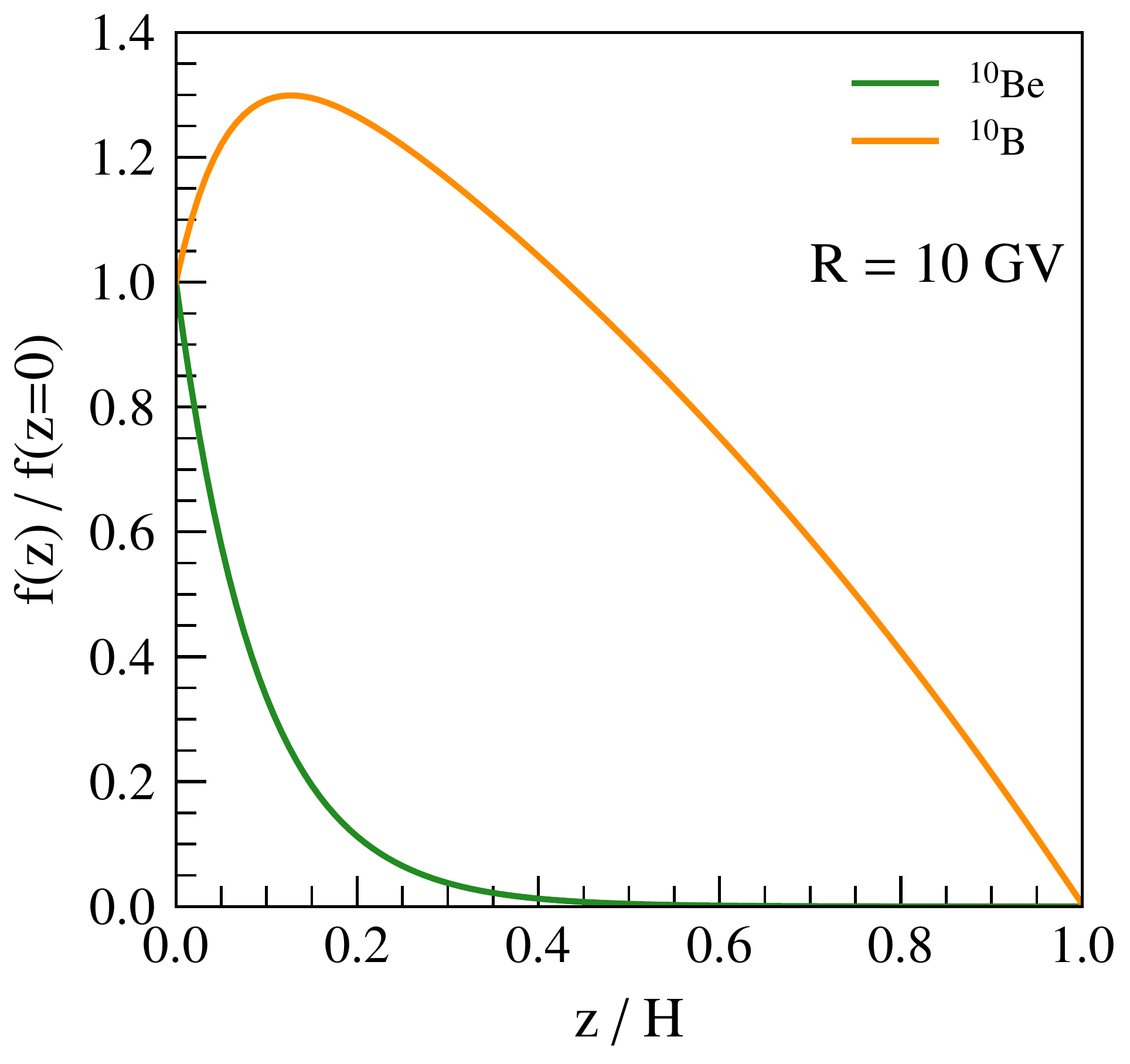}
\end{center}
\caption{Spatial distribution of unstable isotope $^{10}$Be and its daughter $^{10}$B, both normalised to their disc value $f_0(p)$ at rigiditiy 10 GV. In this case $H=6$ kpc has been assumed.}
\label{fig:be10b10-distribution}
\end{figure}

The constants $A$ and $B$ in Eq.~\eqref{eq:stable_halo} are obtained imposing the boundary conditions at the Galactic disc and at the edge of the halo, namely $f_a(p, z=0) = f_{0,a}(p)$ and $f_a(p, z=H) = 0$, to obtain:
\begin{equation} \label{eq:f(z,p)}
f_a(z,p) = f_{0,a}(p) \frac{e^{\alpha_- z}  - e^{\alpha_+ z + (\alpha_- - \alpha_+) H} }{1 - e^{(\alpha_- - \alpha_+) H}} \,.
\end{equation}

The value of the distribution function inside the disc, $f_{0,a}(p)$, can be obtained by integrating Eq.~\eqref{eq:slab} between $0^-$ and $0^+$ which gives
$$
-2D_{a}(p) \left(\frac{\partial f_a}{\partial z}\right)_{z=0^{+}} - 
\frac{2}{3} v_A p \frac{\partial f_{0,a}}{\partial p}
+ $$
$$\frac{\mu v(p) \sigma_{a}}{m} f_{0,a} + \frac{2 h}{p^{2}} \frac{\partial}{\partial p}\left[ p^{2} b_{0,a}(p) f_{0,a}\right] = $$
\begin{equation}
= 2 h q_{0,a}(p) + \sum_{a'>a} \frac{\mu v(p) \sigma_{a' \to a}}{m}f_{0,a'},
\label{eq:f0_1}
\end{equation}

The quantity $D_a \partial f_a/\partial z |_{0^+}$ represents the diffusive flux at the disc position and can be obtained deriving Eq.~\eqref{eq:f(z,p)} with respect to $z$, namely:
\begin{equation}\label{eq:Ddfdx}
\left[ D_a \frac{\partial f_a}{\partial z} \right]_{z=0} = - \frac{v_A}{2} \xi(p) f_{0,a}
\end{equation}
where we have introduced the quantity:
\begin{equation}
\xi(p) = -\frac{(1 - \Delta_a) - (1 + \Delta_a) e^{-v_A \Delta_a H/D_a} }{1-e^{-v_A \Delta_a H/D_a}}   
\end{equation}{}

The meaning of the quantity $\xi$ can be understood applying it to stable elements where $\Delta=1$ and then $\xi= 2/(1-e^{v_A H/D_a})$. In the further limit of diffusion dominated case, i.e. when $D_a \gg v_A H$, we get $\xi\rightarrow 2D_a/(v_A H)$, while in the advection dominated case $\xi\rightarrow 2$. On the other hand, in the case of unstable elements with $\tau_d \ll 4D_a/v_A^2$ we get $\xi \rightarrow \Delta \simeq \sqrt{4D_a/(v_A^2 \tau_d)}$.

That is, we can write the diffusive flux as
\begin{equation} \label{eq:Ddfdx2}
\left[ D_a \frac{\partial f_a}{\partial z} \right]_{z=0} \simeq -f_{0,a} \frac{D_a}{L_a} \,,
\end{equation} 
where $L_a$ represents the maximum propagation distance, namely $L_a=H$ for stable elements in the diffusion dominated limit, $L_a= \infty$ for stable elements in the advection dominated limit and $L_a=\sqrt{D_a \tau_{d,a}}$ for elements decaying on a timescale shorter than $4D_a/v_A^2$.

Following~\cite{Jones2001,Evoli2019}, we rewrite the transport equation in terms of the flux as a function of kinetic energy per nucleon $I_a =A_a p^2 f_{0_a}$:
$$
\frac{I_{a}(E)}{X_{a}(E)} + \frac{d}{dE}\left\{\left[ \left(\frac{dE}{dx}\right)_{ad} +  \left(\frac{dE}{dx}\right)_{{\rm ion},a}\right] I_{a}(E)\right\} + 
$$
\begin{equation}
+ \frac{\sigma_{a} I_{a}(E)}{m} = Q_a(E)
\label{eq:I(E)}
\end{equation}
where
\begin{equation} \label{eq:X(E)}
X(E) = \frac{\mu v}{2 v_A} \, \frac{2 (1-e^{-{v_A \Delta_a H / D_a}})}{(1+\Delta) - (1-\Delta) e^{-{v_A \Delta_a H / D_a}}}
\end{equation} 
is the grammage for nuclei with kinetic energy per nucleon $E$, 
\begin{equation} \label{eq:dEdt_ad}
\left( \frac{dE}{dx} \right)_{\rm ad} = - \frac{2 v_A}{3 \mu c} \, \sqrt{E (E+m_p c^2)}
\end{equation} 
is the rate of adiabatic energy losses due to advection and 
\begin{equation} \label{eq:Q(E)}
Q_a(E) = 2 h \frac{A_{a} p^{2} q_{0,a}(p)}{\mu v} + \sum_{a'>a} \frac{I_{a}(E)}{m}\sigma_{a' \to a},
\end{equation} 
is the source term.
That means that we can adopt the same formal solution of Eq.~\eqref{eq:I(E)} as in~\cite{Evoli2019} for stable species but using for the grammage the expression in Eq.~\eqref{eq:X(E)}.

In Fig.~\ref{fig:be10b10-distribution} we plot the spatial distribution of $^{10}$Be and its decay product $^{10}$B, with the distribution functions both normalised to their value in the disk, computed at fixed rigidity $R=10$ GV and assuming a halo half-thickness $H=6$ kpc. The figure clearly illustrates that the $^{10}$B contributed by the decays of $^{10}$Be is spatially extended, while the $^{10}$B produced through spallation reactions is mainly concentrated in the disc. 

Finally,  it is interesting to study the asymptotic behaviour of Eq.~(\ref{eq:X(E)}) in three different cases: advection-dominated, diffusion-dominated and decay-dominated regimes. The corresponding expressions are:
\[
X = \left.
\begin{cases}
\frac{\mu v}{2 v_A} & \text{when } \tau_{\rm adv} \ll \tau_{\rm diff}\,, \tau_{d} \, \\
\frac{\mu v H}{2 D} & \text{when } \tau_{\rm diff} \ll \tau_{\rm adv}\,, \tau_{d} \, \\
\frac{\mu v}{2} \frac{\tau_d}{\sqrt{ D \tau_d}} & \text{when } \tau_{d} \ll \tau_{\rm diff}\,, \tau_{\rm adv} \,
\end{cases}
\right.
\]

\begin{figure*}[t]
\begin{center}
\includegraphics[width=0.98\columnwidth]{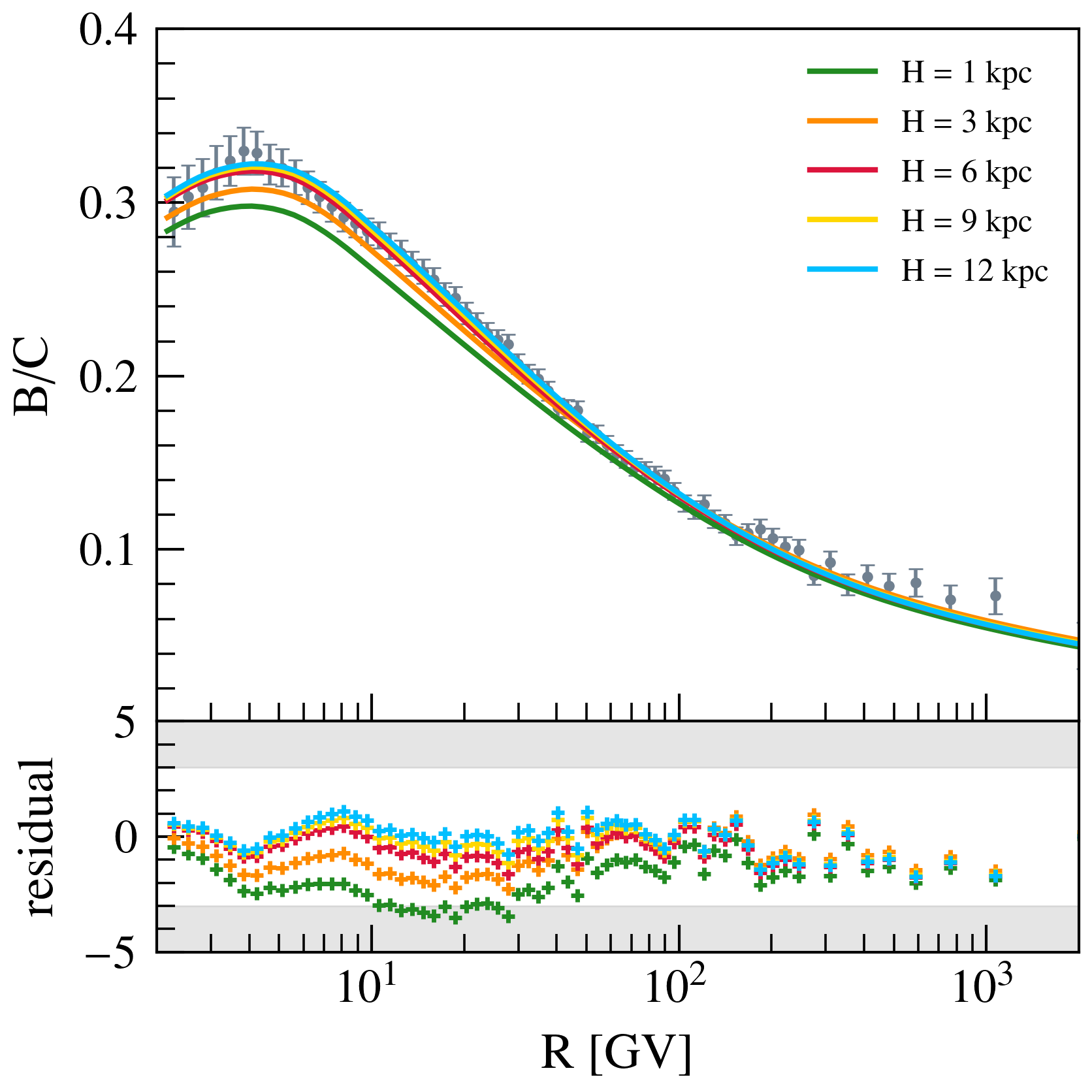}
\hspace{\stretch{1}}
\includegraphics[width=0.98\columnwidth]{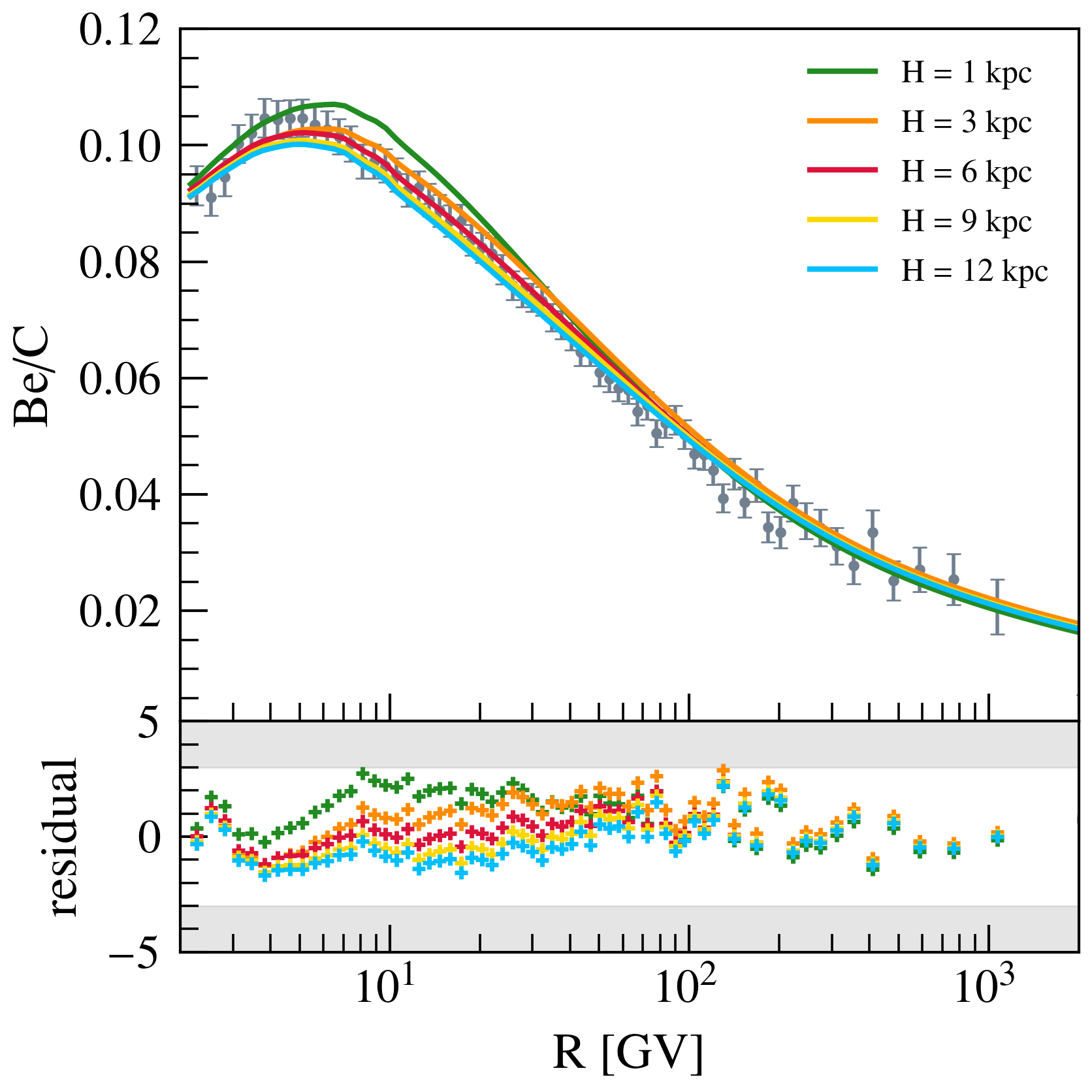}
\end{center}
\caption{Ratio of Boron over Carbon fluxes (left) and Beryllium over Carbon fluxes (right). The data points are the results of measurements by AMS-02 \cite{AMS02-bc} and the error bars are computed with statistical and systematic errors summed in quadrature. The curves illustrate our best-fit results for different values of the halo size $H$. The bottom panels show the corresponding residuals with the same color code.}
\label{fig:bc-bec}
\end{figure*}

This shows how the combination of secondary/primary fluxes which constrain $H/D$ and unstable/stable secondaries which constrain $H/\sqrt{D}$ together allow us to determine both $D$ and $H$ independently, though with all limitations deriving from systematic uncertainties in the experimental data and in the spallation cross sections. 

\subsection{Stable elements with contribution from unstable ones} \label{sec:stable}

When a stable element $a$ receives a contribution from the decay of an unstable element $b$, Eq.~\eqref{eq:halo} becomes
\begin{equation} \label{eq:halo2}
  D_a(p)\frac{\partial^2 f_a}{\partial z^2}
  - v_A \frac{\partial f_a}{\partial z}  
  = - \frac{f_b}{\tau_{d,b}} \, .
\end{equation}

The solution can be obtained with the method of variation of constants:
\begin{multline}\label{eq:fa(z,p)}
f_a(z,p) = \frac{1 - e^{\alpha (z-H)}}{1 - e^{-\alpha H}} \times \\
\times \left\{ f_{a,0} + \frac{1}{\tau_d v_A} \int_{0}^{H} f_{b}(z') \left( 1 - e^{-\alpha z'} \right) dz' \right\} - \\ 
	- \frac{1}{\tau_{d} v_A} \int_{z}^{H} f_{b}(z') 
			\left( 1 - e^{-\alpha (z'-z)} \right) dz'
\end{multline}	
where $\alpha= v_A/D_a$.

In order to get the solution for $f_{a,0}(p)$ at the disk we need to solve again Eq.~\eqref{eq:f0_1} but with a different expression for the diffusive flux term $D_a \partial_z f_a|_{0^+}$ which is  obtained deriving Eq.~\eqref{eq:fa(z,p)} with respect to $z$. The result is easily found to be:
\begin{equation}
  D_a \left. \frac{\partial f_{a}}{\partial z} \right|_{0^+}
     = - \frac{v_A f_{a,0}}{e^{\alpha H} - 1}
     + \frac{1}{\tau_{d,b}}  \int_{0}^{H} f_{b}(z')  \frac{e^{\alpha (H-z')} - 1}{e^{\alpha H} - 1} dz' \,.
\end{equation}

This result can be further simplified using Eq.~\eqref{eq:f(z,p)} for the spatial dependence of the distribution $f_b(z)$ of the radioactive CR  and performing the integral. The final expression can be explicitly written and reads:
\begin{multline} \label{eq:Dflux_a}
  D_a \left. \frac{\partial f_{a}}{\partial z} \right|_{0^+}
     = - \frac{f_{a,0}(p) v_A}{e^{v_A  H / D_a} - 1} 
       + \\ f_{b,0}(p) v_A \left[ \Delta_b \coth\left( \frac{v_A H \Delta_b}{2 D_a}\right) 
       	- \coth\left( \frac{v_A H}{2 D_a}\right) 
	\right]  \,.
\end{multline}

The second term $\propto f_{b,0}$ represents an effective injection due to the decay of the species $b$. This term disappears when $\tau_{d,b}\rightarrow \infty$ since $\Delta_b \rightarrow 1$. 

When Eq.~\eqref{eq:Dflux_a} is plugged into Eq.~\eqref{eq:f0_1} we get a formal solution for $f_{a,0}(p)$ identical to Eq.~\eqref{eq:I(E)} but with a different injection term, which is now the sum of the secondary source term and the source term due to the Be decay, i.e.
\begin{multline} \label{eq:Q0_a}
Q_{0,a}(p) = 2 h q_{0,a}(p) + 2 f_{b,0}(p) v_A \left[ \Delta_b \coth\left( \frac{v_A H \Delta_b}{2 D_a}\right) \right. - \\
\left. \coth\left( \frac{v_A H}{2 D_a}\right) \right] \equiv 2 h q_{0,a}(p) + 2 \tilde{q}_{b \rightarrow a}(p)
\end{multline}

\begin{figure*}[t]
\begin{center}
\includegraphics[width=0.99\columnwidth]{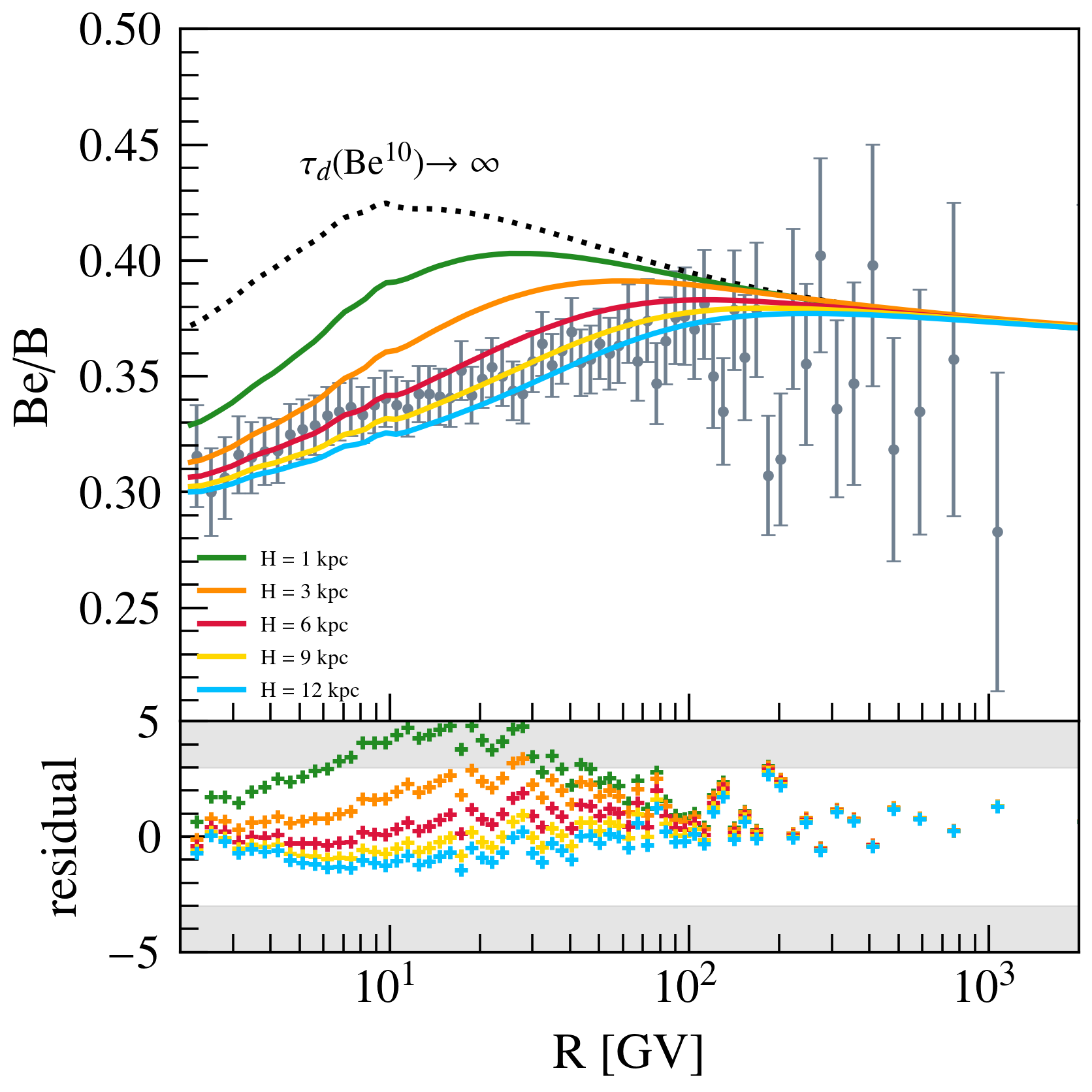}
\hspace{\stretch{1}}
\includegraphics[width=0.99\columnwidth]{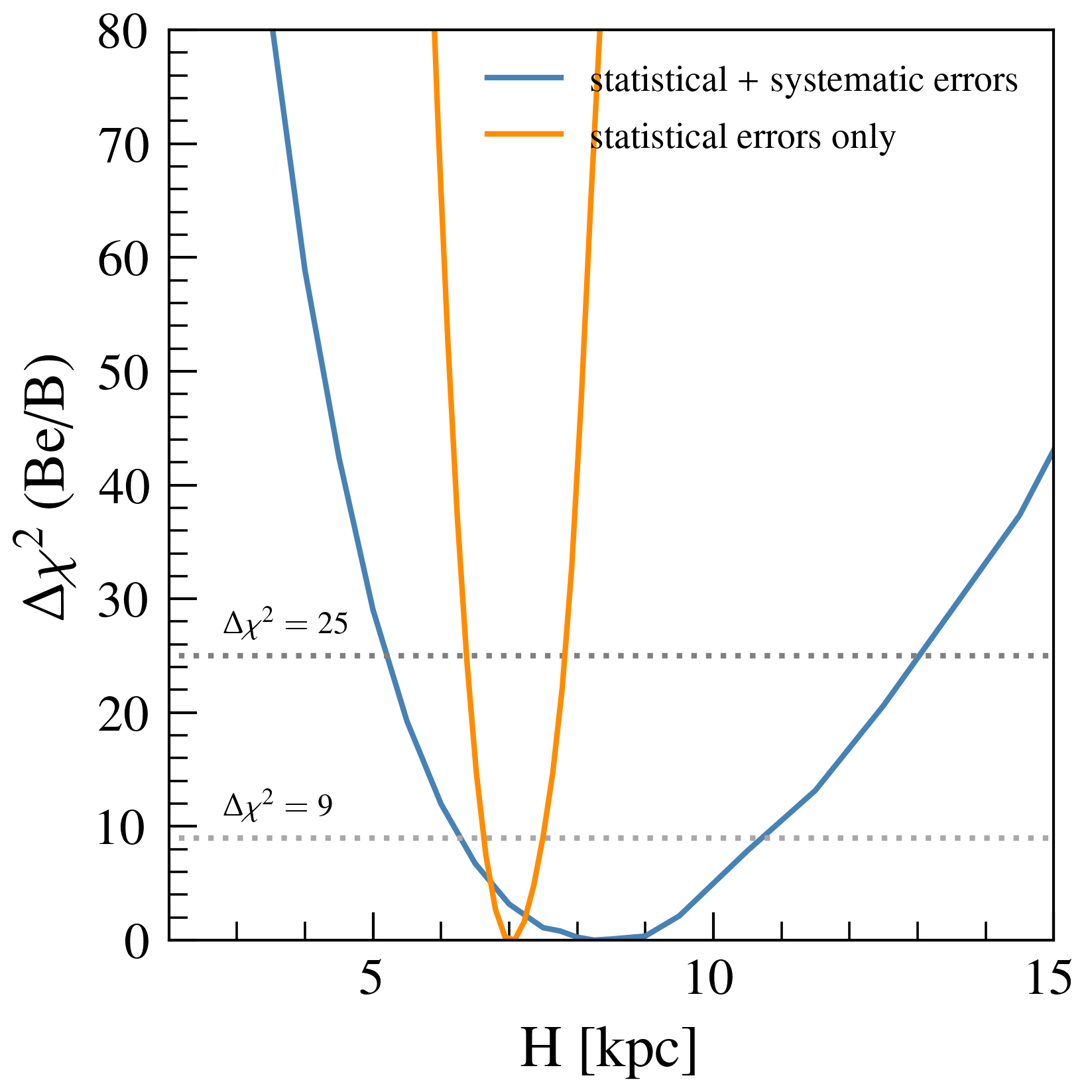}
\end{center}
\caption{{\it Left panel:} Ratio of Beryllium over Boron fluxes. The dotted line shows the case without decay for $^{10}$Be while the other lines refer to different values of $H$, as labeled. {\it Right panel:} $\Delta \chi^2 \equiv \chi^2 - \chi^2_{\rm min}$ computed on the Be/B data as a function of the halo size $H$. We show both the case where only the statistical errors are used (solid orange) and the case with the total errors (solid blue). The best-fit reduced $\chi^2$'s are $\sim$3 and $\sim$0.85 in the two cases. The allowed maximum  $\chi^2$ at $3\sigma$ and $5\sigma$ are also indicated with dotted lines.}
\label{fig:beb-chi2}
\end{figure*}

\section{Results}
\label{sec:Results}

\subsection{Secondary over primary ratios}

In this section we present our results obtained through a single multivariate fitting procedure to compare AMS-02 experimental data with theoretical spectra computed as discussed in the previous sections. For each value of the halo half-thickness $H$ we minimise the $\chi^2$ with respect to the AMS-02 data on Be/C, B/C, Be/O and B/O~\cite{AMS02-libeb} and C, N, O~\cite{AMS02-heco,AMS02-nitrogen}, the latter data limited to rigidities larger than 10 GV, so as to make the results only weakly dependent upon the uncertainties typical of the low energies.

The set of parameters varied along the minimising procedure are: solar modulation potential $\phi$, advection velocity $v_A$, diffusion coefficient constants $D_0$ and $\delta$, {injection power law index $\gamma_{{\rm inj}}$ (assumed to be the same for all the primary species) and injection efficiency $\epsilon_a$, the latter quantity being species dependent.}

As in \cite{Evoli2019}, the spallation network is computed starting from iron (Z = 26) all the way down to Lithium. The injection efficiency for nuclei heavier than oxygen, where AMS-02 data are not yet available, are fitted against the high-energy CREAM data~\cite{Ahn2009}.

The combined fit of the ratios Be/C, B/C, Be/O and B/O constrains $\delta \sim 0.54$, for any value of $H$ in the range $1-20$ kpc. Coherently, the injected slope is fitted to be $\gamma_{\rm inj} \sim 4.3$, the ratio $D_0/H$ is $\sim 0.44$ (in units of $10^{28}$ cm$^2$ s$^{-1}$ kpc$^{-1}$), $\phi = 0.68$~GV and $v_A \sim 5$~km/s.
We notice therefore that the typical dependence of the B/C ratio with respect to the quantity $D_0/H$ is maintained also if the radioactive decay of $^{10}$Be is taken into account, although the $\chi^2$ associated with different values of $H$ is not the same. In particular, the $\chi^2$ appears to be higher for smaller values of $H$.

In Fig.~\ref{fig:bc-bec} we show the comparison of our best-fit results with the AMS-02 data on the ratios B/C (left panel) and Be/C (right panel) for different values of $H$ as labeled. In these plots we show the total experimental uncertainty, obtained summing in quadrature the statistical and systematic errors as published by the AMS-02 Collaboration \cite{AMS02-bc,AMS02-heco,AMS02-libeb}. As expected, for low values of $H$, say $\sim 1$ kpc, the effect of $^{10}$Be decay is weak, thereby leading to overestimating the Be/C ratio and underestimating the B/C ratio. 

In Fig.~\ref{fig:bc-bec}, as in the forthcoming figures, we plot also the residual respect to experimental data, defined as the "distance" between the theoretical expectation and data divided by the total experimental error. As follows from Fig.~\ref{fig:bc-bec}, the residual is always confined within 3$\sigma$, confirming a good accuracy of our fitting procedure. 

The residuals clearly show a preference for relatively large values of the halo size, $H \gtrsim 6$ kpc. A similar conclusion can be drawn by considering the Be/O and B/O ratios, not shown here. A quantitative assessment of the significance of these fits will be discussed in Section~\ref{sec:BeB} using the Beryllium over Boron ratio.

\subsection{Beryllium over Boron ratio}\label{sec:BeB}

In order to calculate the Be/B ratio, we solve the transport equations for all isotopes of both beryllium ($^{7}$Be, $^{9}$Be and $^{10}$Be) and boron ($^{10}$B and $^{11}$B). As we discuss below, this ratio is more sensitive to the value of $H$ with respect to the secondary to primary ratios. 

If all isotopes of Be were stable, the Be/B ratio at rigidities above $\sim 10$ GV would be a slowly decreasing function of energy, up to about $\sim 200$ GV, where the spallation time of Be becomes appreciably longer than the escape time from the Galaxy. The slight decrease reflects the fact that the total inelastic cross section scales as  $\propto A^{0.7}$ and boron (denominator) is slightly heavier than beryllium. At higher rigidity, since the production cross sections are basically independent of energy~\cite{Evoli2019}, the Be/B ratio is expected to be constant. 
Moreover, the spallation of Boron increases the amount of Beryllium (numerator) at the same energy per nucleon. 
This behaviour is shown as a black dotted line in the left panel of Fig.~\ref{fig:beb-chi2}. At rigidities $\lesssim 10$ GV the spallation cross section acquires a small energy dependence which reflects in the small increase with rigidity visible in the figure. 

\begin{figure}[t]
\begin{center}
\includegraphics[width=\columnwidth]{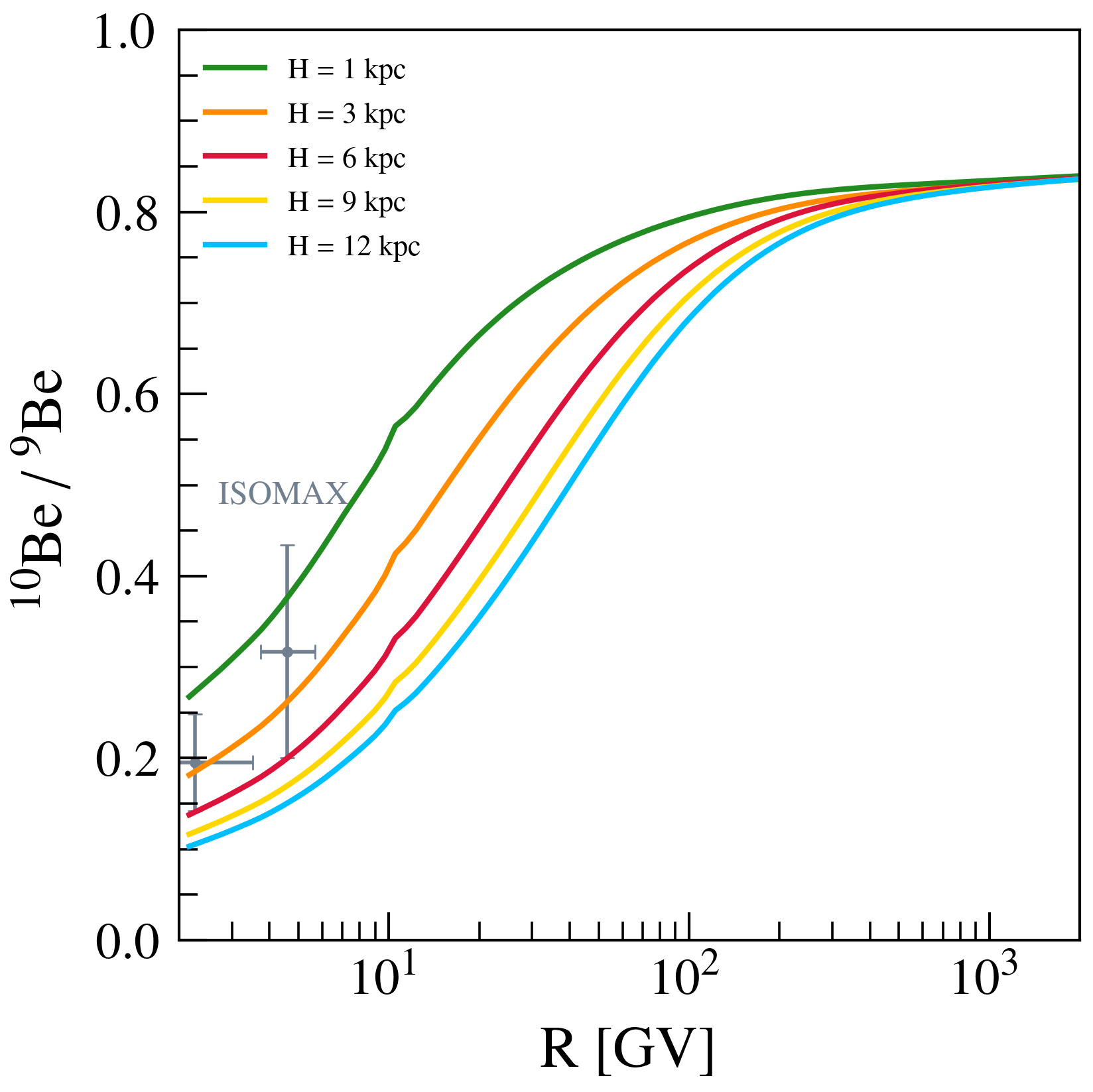}
\end{center}
\caption{Predicted $^{10}$Be/$^9$Be ratio for different values of the halo size $H$. Data points are from the ISOMAX experiment~\cite{Hams2004}.}
\label{fig:be10be9}
\end{figure}

The AMS-02 data clearly show that the Be/B ratio increases with rigidity at least up to $\sim 100$ GV. The simplest explanation of such a trend is based on the decay of $^{10}$Be at low rigidity, where decays occur faster than escape. The coloured solid lines in the left panel of Fig.~\ref{fig:beb-chi2} show the results of our calculations for the best-fit to the secondary-over-primary ratios for different values of $H$ as found in the previous section.

The residuals are also shown in the bottom part of the left panel of Fig.~\ref{fig:beb-chi2}. 
In the right panel of the same Figure we plot as a function of $H$ the $\Delta \chi^2$ (defined as the difference between the $\chi^2(H)$ and its minimum $\chi^2_{\rm min}$) computed on the Be/B data and assuming statistical and systematic errors summed in quadrature (blue solid line) from which one can infer that values of $H \lesssim 6$~kpc are disfavoured at more than $99.73\%$ of confidence (3$\sigma$), while $H \lesssim 5$~kpc appears to be excluded at more than $5\sigma$. These two C.L.'s correspond to $\Delta \chi^2 =9 $ and $25$ respectively.
This result is in agreement with the estimates based on the comparison between numerical models for the CR electron distribution and the morphology of the diffuse radio emission~\cite{Orlando2013,DiBernardo2013}.

It might be argued that the $\chi^2$ of the fit has a well-defined statistical significance only with respect to statistical errors, although systematics (for instance in the energy determination, but not only) can change the number of events that belong in a given rigidity bin. In the right panel of Fig.~\ref{fig:beb-chi2} {we also show the reduced $\chi^2$ as a function of $H$ as calculated with respect to the statistical errors only.} 
Clearly, the predictive power of the former case is higher than the latter, although the statistical significance gets smaller because of the very small statistical error bars of AMS-02 data. Nevertheless, it leads to an allowed range for $H$ from $\sim$6.5 kpc to $\sim$7.5 kpc at 99.7\% of confindence level, with a best fit of $H \sim 7$ kpc, thereby confirming the previous finding based on B/C and Be/C ratios. 

As pointed out in Sec.~\ref{sec:Model}, the weighted slab model adopted here is not suitable to describe the transport of unstable isotopes when the decay takes place inside the thickness of the Galactic disc. This restricts the range of applicability of our calculations to rigidities $\gtrsim$ few GV. On the other hand, existing measurements of the $^{10}$Be/$^{9}$Be ratio \cite{Webber1979,Garcia-Munoz1981,Connell1998,Hams2004,Lukasiak1999} are limited to sub-GV rigidities. In the near future, the HELIX (High Energy Light Isotope eXperiment) mission~\cite{Park2019} aims at measuring this ratio up to tens of GV. For the sole purpose of illustrating the capabilities required of future experiments in order to discriminate among different values of $H$, in Fig.~\ref{fig:be10be9} we plot the expected $^{10}$Be/$^9$Be ratio for different values of $H$, compared with data points from ISOMAX \cite{Hams2004} that collected data reaching up to few GV rigidity. 

\begin{figure}[t]
\begin{center}
\includegraphics[width=\columnwidth]{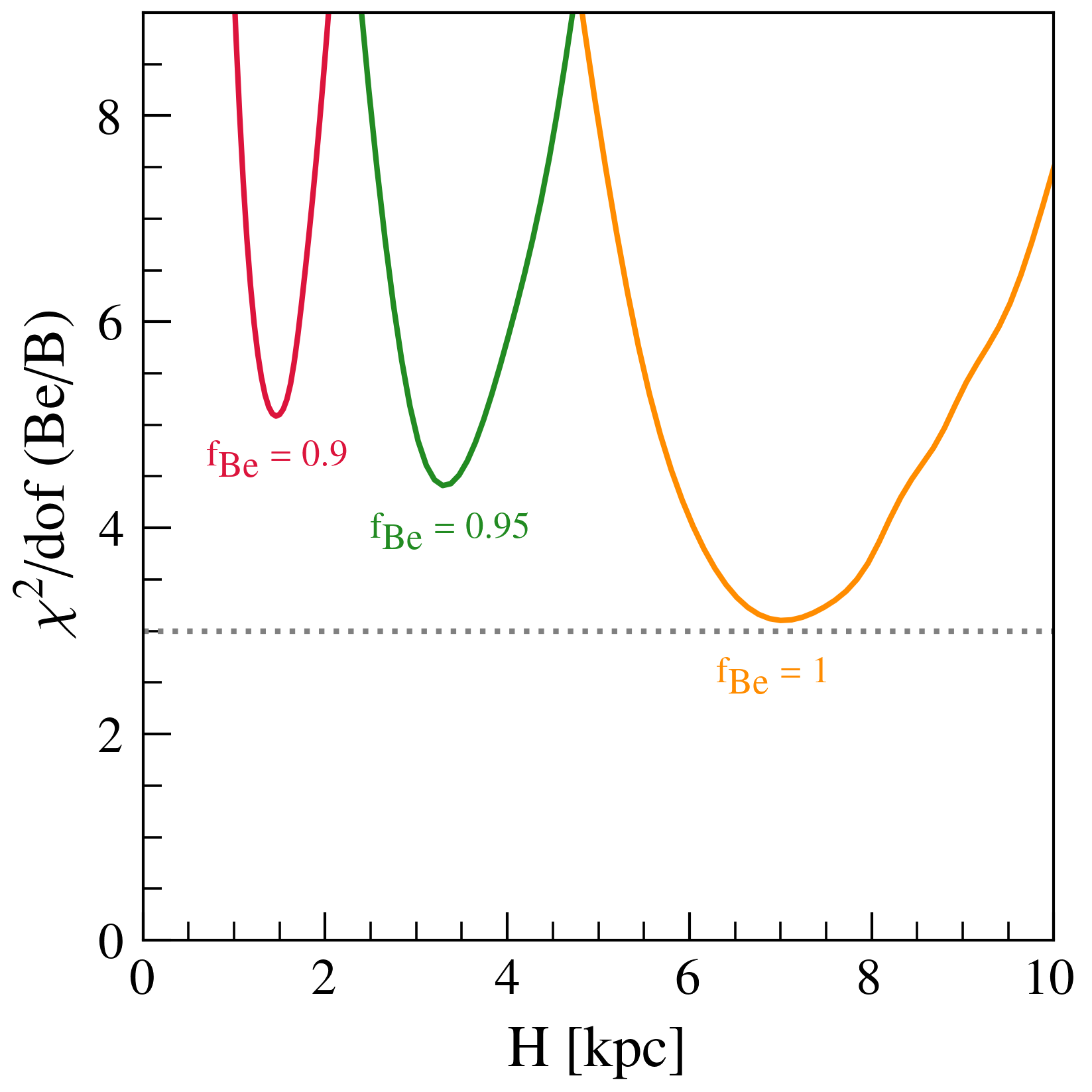}
\end{center}
\caption{The reduced $\chi^2$ indicator, calculated using statistical errors only, is shown as a function of the halo size $H$ for different values of the normalization factor for the Be production cross-sections $f_{\rm Be}$.}
\label{fig:cross}
\end{figure}

We used as a benchmark case the one corresponding to $H=6$ kpc and asked the following question: how good a measurement a future experiment should perform in order to measure $H$ within a given accuracy? From Fig.~\ref{fig:be10be9} we infer that an accuracy better than 30$\%$ in the measurement of the $^{10}$Be/$^9$Be ratio is needed in order to allow us to discriminate between $H=3$ and $H=6$ kpc. An accuracy better than 10$\%$ is necessary to distinguish between $H=6$ and $H=9$ kpc. This level of accuracy is expected to be within reach for the HELIX mission \cite{Park2019}.

\subsection{Effects of the uncertainties in the spallation cross sections}

As discussed in detail in Refs.~\cite{Genolini2015,Tomassetti2015,Evoli2019}, the main limitation in extracting physical information on CR transport from secondary to primary ratios derives from uncertainties in the spallation cross sections. The same limitations holds for the Be/B ratio, to an extent that we describe below. 

As discussed in \cite{Evoli2019}, although the energy dependence of spallation cross sections is known to be weak, their normalization is uncertain by factors that, depending on the nucleus, can be tens of percent to order unity. Here we parametrize the uncertainty in the production of beryllium in terms of a fudge factor $f_{\rm Be}$, while assuming that boron production is known. In other words, this fudge factor can be interpreted as a relative uncertainty between the production of beryllium and that of boron. 

A quick inspection of Fig.~\ref{fig:beb-chi2} leads to some qualitative conclusions: a decrease in the parameter $f_{\rm Be}$ causes the horizontal trend at high rigidity to get lower, thereby making the low energy part closer to the AMS-02 data for smaller values of $H$. On the contrary, an increase in $f_{\rm Be}$ leads to a worse fit in general. This trend is confirmed quantitatively in Fig.~\ref{fig:cross}, where we show the reduced $\chi^2$ of the ratio Be/B, calculated taking into account only statistical errors, as a function of the halo size $H$ and for different values of the fudge factor $f_{\rm Be}$.
The best $\chi^2$ is still the one obtained for $f_{\rm Be}\sim 1$ and $H\sim 7$ kpc. However, a reduction of the cross section of beryllium production by $10\%$ would imply a halo size $H\sim 1.5$ kpc, although with a worse $\chi^2$ with a difference between the two situations of $\Delta \chi^2 \sim 2$.

This conclusion illustrates in a clear way the importance of having reliable measurements of the spallation cross sections, as already pointed out in Ref.~\cite{Evoli2019} based on the secondary to primary ratios. 

\section{Conclusions}
\label{sec:Conclusions} 

We are going through a very peculiar time in the history of the investigation of the origin of CRs: on one hand the AMS-02 data have projected us into a precision era of the measurement of CR fluxes, that in principle should allow to solve some long lasting problems in the field. For instance such data have allowed us to detect features in the spectra of primaries that most likely are telling us about scattering properties of CRs in their journey through the Galaxy and the measurement of the secondary to primary ratios have provided the best measurement of the grammage traversed by CRs. 

On the other hand, some pieces of these measurements, such as the spectrum of positrons and antiprotons, have opened a huge space for models of CR transport that seem to question the very bases on which the points listed above are based. To be more precise, there have been claims that the observed trends in the positron and antiproton fluxes may suggest that CRs accumulate most of the grammage in regions near their sources rather than on Galactic scales~\cite{Cowsik2016,Lipari2017}. These models make some clear predictions: 1) primary electrons should be injected with a spectrum that is different from that of primary nuclei; 2) positrons and antiprotons are solely secondary products of hadronic CR interactions; 3) the confinement time of CRs in the Galaxy must be weakly dependent or independent upon energy and much shorter that the loss timescales for positrons, at least for $E\lesssim$~TeV. This implies that the observed spectrum of $e^{-}$ is roughly the same as that at the source and the spectrum of $e^{+}$ is approximately the same as that of their parent protons, at least up to $E\lesssim$~TeV. The recent results of DAMPE~\citep{DAMPE-leptons} for the $e^{+}+e^{-}$ spectrum is usually cited as a possible proof in support of this scenario. 

The critical assumption in these models is that the confinement time of CRs in the Galaxy is much shorter than believed. The most sensitive measurement of the confinement time is provided by the abundance of $^{10}$Be, the unstable isotope of beryllium, compared with the abundance of the stable isotopes. Unfortunately the $^{10}$Be/$^{9}$Be ratio has only been measured at low rigidities where predictions are extremely model dependent because of the fact that the radioactive decays occur inside the thin Galactic disc, where diffusion processes, advection and the local structure of the magnetic field are, to say the least, poorly known. 

The recent measurement of the Be/B ratio performed by AMS-02 and extended up to $\sim$TV rigidities has given us the opportunity to reconsider the issue of the confinement time: the decay of the unstable $^{10}$Be makes the ratio acquire a peculiar increase with rigidity at rigidity $\lesssim 100$~GV that carries information about the time scale of CR transport, and more specifically about the size of the Galactic halo $H$. We have shown that this increase is best reproduced assuming $H$ of $\sim 7$ kpc, which corresponds to a transport time in the Galaxy that is incompatible with the assumption of loss free propagation of positrons. The minimum value of $H$ that appears to be compatible with the measured Be/B ratio is $H\sim 5$ kpc.

There are two caveats that affect this conclusion and need to be discussed.

First, the conclusions above are derived using the spallation cross section taken at face value, as given by the best available fits~\cite{Evoli2019}. However, allowing for a $\sim 10\%$ uncertainty in the production cross section of beryllium nuclei makes the conclusion more shaky, in the sense that the value $H \sim 7$ kpc remains the most likely value of the halo size, but the reduced $\chi^{2}$ for $H=1.5$ kpc is only $\sim 2$ points larger than in the previous case, if the cross section is $\sim 10\%$ smaller than expected. We confirm the urgent need for a campaign of high precision measurements of the spallation cross sections (e.g., an experimental program to measure carbon-proton and oxygen-proton cross sections at 13 A GeV/c at the CERN SPS has been proposed in~\cite{Aduszkiewicz:2287004}), so as to allow us to finally comprehend the CR transport in its subtle features. 

Second, the weighted-slab model adapted here is strictly valid for halo size smaller than the gradients in the radial direction. Since our best-fit $H \sim 7$ kpc is comparable with the distance to the Galactic Centre a more reliable estimate should be derived using three-dimensional transport models.

In Sec.~\ref{sec:BeB} we made some predictions of the performance required of future experiments, such as HELIX~\cite{Park2019} in order to discriminate among different values of $H$. An accuracy better than 30$\%$ in the measurement of the $^{10}$Be/$^9$Be ratio is needed in order to allow us to discriminate between $H=3$ and $H=6$ kpc, while an accuracy better than 10$\%$ is necessary to distinguish between $H=6$ and $H=9$ kpc. This requirements should be fulfilled by the HELIX mission. 

\begin{acknowledgments}
C.E.~acknowledges the European Commission for support under the H2020-MSCA-IF-2016 action, Grant No.~751311 GRAPES 8211 Galactic cosmic RAy Propagation: An Extensive Study. This work was partially funded through Grants ASI/INAF No.~2017-14-H.O
\end{acknowledgments}

\bibliographystyle{myapsrev4-2}
\bibliography{crams} 

\end{document}